\documentclass[letterpaper, 10 pt, conference]{ieeeconf}  
\usepackage{amsmath,graphicx}
\usepackage{amsfonts}
\usepackage[colorlinks,linkcolor=blue]{hyperref}
\usepackage{array}
\IEEEoverridecommandlockouts                              

\title{\LARGE \bf
Photoacoustic Digital Skin: Generation and Simulation of Human Skin Vascular for Quantitative Image Analysis
}

\author{Tengbo Lyu, Changchun Yang, Jiadong Zhang, Shanshan Guo, Feng Gao, and Fei Gao$^{*}$
\thanks{Tengbo Lyu, Changchun Yang, Jiadong Zhang, Shanshan Guo, Feng Gao and Fei Gao are with the Hybrid Imaging System Laboratory, Shanghai Engineering Research Center of Intelligent Vision and Imaging, School of Information Science and Technology, ShanghaiTech University, Shanghai 201210, China (*corresponding author: gaofei@shanghaitech.edu.cn).}
}

\begin{document}

\maketitle
\thispagestyle{empty}
\pagestyle{empty}

\begin{abstract}

Photoacoustic computed tomography (PACT) is a hybrid imaging modality, which combines the high optical contrast of pure optical imaging and the high penetration depth of ultrasound imaging. 
However, photoacoustic image dataset with good quality and large quantity is lacking. In this paper, we mainly talk about how to generate a practical photoacoustic dataset. 
Firstly, we extracted 389 3D vessel volumes from CT whole-lung scan database, and enhanced the blood vessel structures. 
Then for each 3D vessel volume, we embedded it into a three-layer cubic phantom to form a skin tissue model, which includes epidermis, dermis, and hypodermis.
The vessel volume was placed randomly in dermis layer in 10 different ways. Thus, 3890 3D skin tissue phantoms were generated. 
Then we assigned optical properties for the four kinds of tissue types. 
Monte-Carlo optical simulations were deployed to obtain the optical fluence distribution. 
Then acoustic propagation simulations were deployed in k-Wave toolbox to obtain the photoacoustic initial pressure. 
Universal back-projection algorithm was used to reconstruct the photoacoustic images. This dataset could be used for deep learning-based photoacoustic image reconstruction, classification, registration, quantitative image analysis. This dataset could be accessesed on the website: \href{https://dx.doi.org/10.21227/h8hw-6a03}{https://dx.doi.org/10.21227/h8hw-6a03}.
\newline

\indent \textit{Clinical relevance}— This paper mainly contributes to establish a giant 3D skin vascular phantom dataset which contains 3890 skin tissue models. This dataset could be used for deep learning-based blood oxygenation quantification and other tasks such as photoacoustic reconstruction and classification study.
\end{abstract}

\section{INTRODUCTION}

Photoacoustic (PA) tomography (PAT) has been widely studied during the past two decades. PAT combines both the advantages of high optical contrast and deep ultrasound penetration depth, which makes it a good choice for quantitative imaging in deep tissue \cite{zhou2016tutorial}. 
There are two main PAT-based imaging modalities, which are photoacoustic computed tomography (PACT) and photoacoustic microscopy (PAM) \cite{wang2008tutorial}. 
PACT reconstructs the target object image in a diffusive regime with the wide-field illumination, and then uses an ultrasound transducer array to detect the induced acoustic signals at multiple different locations. 
The geometrical shape of the transducer could be set as line, circular, spherical, or cylindrical shapes according to the 2D or 3D imaging setup. 
After obtaining the raw data of the PA signals, the next step is to reconstruct the original images using reconstruction algorithms such as commonly used delay and sum, time reversal, and universal back-projection algorithms.
While PAM mainly images the object targets in a scanning manner using focused pulse light or a focused single-element ultrasound transducer, which provides higher spatial resolution with limited imaging depth. 
In 3D PA image reconstruction, the amplitude of a voxel in a PA image can be obtained by 
\begin{equation}
{p_0}\left( {{\bf{x}},\lambda } \right) = {\mu _a}\left( {{\bf{x}},\lambda } \right)\Gamma \left( {\bf{x}} \right)\Phi \left( {{\bf{x}},\lambda ;{\mu _a},s,g} \right)
\end{equation}
where ${p_0}$ is the PA initial pressure, ${\bf{x}}$ is the voxel’s location within the sample, $\lambda $ is the optical wavelength, ${\mu _a}$ is the optical absorption coefficient, $g$ is the optical anisotropy factor,$\Gamma $ is the PA efficiency and $\Phi $ is the optical fluence.

Deep learning-based tasks such as image recognition, classification, registration, and reconstruction are data dependent, especially large-scale database \cite{yang2020deep}. 
However, in PA image analysis, there is not enough large-scale image dataset released. 
What’s more, 3D PA image dataset is also lacking. With this motivation in mind, in this paper, we built a giant 3D skin tissue phantom set, which includes 3890 phantoms. 
After assigning different optical properties for each tissue, we deployed optical and acoustic simulations under 16 wavelengths, which range from 750 nm to 900 nm with step of 10 nm. 
This giant 3D skin tissue model dataset as well as the optical and acoustic simulation images could be accessed on the website: \href{https://dx.doi.org/10.21227/h8hw-6a03}{https://dx.doi.org/10.21227/h8hw-6a03}.

In the following several sections, we will cover how to generate the 3890 3D skin tissue phantoms in Section II, how Monte-Carlo based optical simulation is run in Section III and Section IV. A brief conclusion will be shown in Section V.

\section{3D SKIN TISSUE PHANTOM GENERATION}
Humans’ skin tissue is composed of three layers, which are epidermis, dermis, and hypodermis layer, respectively. 
The main optical absorption in the epidermis layer is contributed by melanosomes. In the dermis layer, the main optical absorption is from blood. 
In the hypodermis layer, the main optical absorption is contributed by lipid and fat, where optical absorption by blood could be neglected \cite{jacques2013optical}. Based on these facts, we generate our 3D skin tissue phantoms by assigning three different layers to the number 1, 2, 3, which stands for the epidermis, dermis, and hypodermis layer, respectively (shown in Fig. 1).
\begin{figure}[!t]
\centering
\includegraphics[width=3.3in]{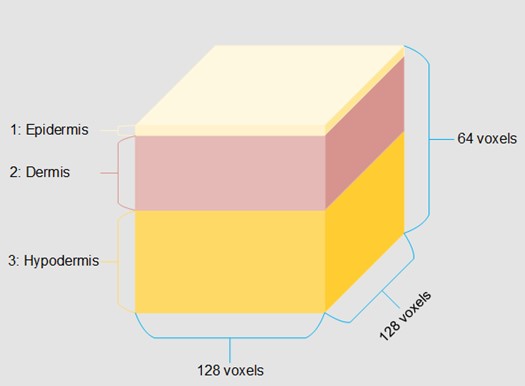}
\caption{A simple illustration of the 3D skin structure phantom model.}
\end{figure}

The thickness along the Z direction of the epidermis layer is assigned by a random number ranging from 1 to 5 voxels. 
The thickness of the dermis layer is assigned by a random number ranging from 31 to 40 voxels. Since the total thickness of the 3D skin tissue model is set to 64 voxels, the thickness of the hypodermis layer is determined by subtracting the total thickness of the epidermis layer and dermis layer. 
The next step is to embed the blood vessel to the dermis layer.

Yang et al. generated the blood vessel volumes by randomly generating a few straight cylinders filled with blood \cite{yang2019quantitative}. 
Although the number and direction of cylindrical blood vessel is different from each other in their work, it is still too far from the real scenarios. Therefore, in this work, we choose to segment a large number of human lung blood vessel volume from ELCAP Public Lung Image Database (\href{http://www.via.cornell.edu/lungdb.html}{http://www.via.cornell.edu/lungdb.html}). 
This database currently consists of an image set including 50 patients’ low-dose documented whole-lung 3D CT scans. 
The X direction and Y direction are both 512 voxels. 
While along the Z direction, the pixel size varies from 212 to 304 voxels. 
Since we only focus on the blood vessel structure and ignore the other physiological structures of the lungs, we need to firstly extract the blood vessel part. 
Thus, we preprocessed the whole-lung CT scan by enhancing the blood vessel using a 3D Frangi filter \cite{frangi1998multiscale}.

For every patient’s scan, we segmented four blood vessel volumes from the left lung and the right lung, respectively. 
Thus, 400 smaller vessel volumes can be obtained. 
However, after we checked the 400 vessel volumes one by one, we found that some vessel volumes lack enough vessel structure. 
As a result, we deleted 11 vessel volumes from the 400 vessel volumes, leaving 389 vessel volumes embedded in the dermis layer. 
The size of each 3D vessel volume is 90, 90, and 30 voxels. 
Before generating the 3D skin tissue models, some image preprocessing was performed to remove disturbing noise-like structure (shown in Fig. 2). 
Then for every blood vessel volume, we randomly flipped them upside down, left and right to make our 3D skin tissue model dataset bigger. 
The random number was set to 10, so finally we got 3890 different blood vessel volumes. 
The number of the blood vessel was set to 4.

To reduce the running time of subsequent optical and acoustic simulations, the X-Y transversal surface of our dataset was set to 128 by 128 pixels. 
Thus, the size of the complete 3D skin tissue model is 128 by 128 by 64 voxels.

The last but very important step was to determine the optical absorption properties of epidermis, dermis, hypodermis and blood vessel. 
According to the work of Jacques et al \cite{jacques2013optical}, we calculated these optical absorption properties of the four tissue types. 
Specifically, the main optical properties of our 3D skin tissue model are optical absorption coefficients, reduced scattering coefficient, refractive index, and anisotropy, which could be represented respectively as ${\mu _a}$(cm$^{-1}$), ${\mu _s}$(cm$^{-1}$), $g$ and $n$. 

\begin{figure}[!t]
\centering
\includegraphics[width=3.5in]{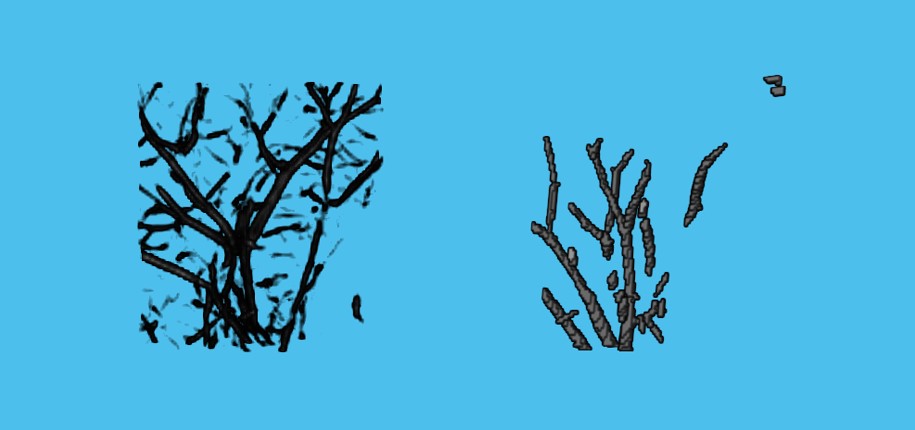}
\caption{The segmented 3D vessel volume (left) and its respective enhanced result (right).}
\end{figure}

\begin{figure}[!t]
\centering
\includegraphics[width=3.5in]{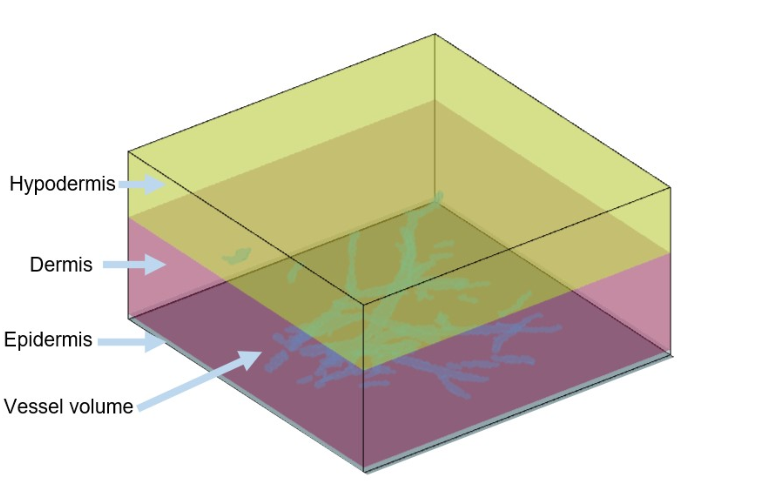}
\caption{One of the 3890 generated 3D skin tissue models structure which consists of 4 kinds of tissue types.}
\end{figure}

\renewcommand\arraystretch{1.5}
\begin{table*}[h]
\caption{Optical properties of the 3D skin tissue phantoms \cite{jacques2013optical}, \cite{bench2020towards}}
\begin{center}
\setlength{\tabcolsep}{6.6mm}{
\begin{tabular}{ccc}
\hline
Tissue type & Parameters & Values\\
\hline
 & Melanosome fraction ${C_M}$ & 1.3-6.3\%, 11-16\%, and 18-43\% for Caucasian, xanthoderma and pigmented skin\\
  & Optical absorption (cm$^{-1}$) & ${\mu _{ae}} = 6.6 \times {10^{11}}{C_M}{\lambda ^{ - 3.33}} + \left( {1 - {C_M}} \right)\left( {0.244 + 85.3{e^{ - \frac{{\lambda  - 154}}{{66.2}}}}} \right)$ \\
 Epidermis &Reduced scattering (cm$^{-1}$) & ${\mu _s} = 68.7{\left( {\frac{\lambda }{{500}}} \right)^{ - 1.16}}$ \\
 &Refractive index & A random number between 1.42 and 1.44 \\
 & Anisotropy & A random number between 0.8 and 0.95 \\
\hline
& Optical absorption (cm$^{-1}$) & ${\mu _{ab}} = {C_{Hb}}{\alpha _{Hb}} + {C_{Hb{O_2}}}{\alpha _{Hb{O_2}}}$ \\
Blood & Reduced scattering (cm$^{-1}$) & ${\mu _s} = 22{\left( {\frac{\lambda }{{500}}} \right)^{ - 0.66}}$ \\
& Refractive index & 1.36 \\
& Anisotropy & 0.994 \\
\hline
& Blood volume fraction ${C_B}$ & 0.2-7\% \\
& Optical absorption (cm$^{-1}$) & ${\mu _{ad}} = {C_B}{\mu _{ab}} + \left( {1 - {C_B}} \right)\left( {0.244 + 85.3{e^{ - \frac{{\lambda  - 154}}{{66.2}}}}} \right)$ \\
Dermis & Reduced scattering (cm$^{-1}$)) & ${\mu _s} = 45.3{\left( {\frac{\lambda }{{500}}} \right)^{ - 1.292}}$ \\
& Refractive index & $n = A + \frac{B}{{{\lambda ^2}}} + \frac{C}{{{\lambda ^4}}}$, $A = 1.3696$, $B = 3916.8$, $C = 2558.8$ \\
& Anisotropy & A random number between 0.8 and 0.95 \\
\hline
& Optical absorption (cm$^{-1}$) & ${\mu _{ah}} =  - \frac{1}{{600}}\lambda  + \frac{{143}}{{60}}$ \\
Hypodermis & Reduced scattering (cm$^{-1}$)) & ${\mu _s} =  - \frac{{11}}{{600}}\lambda  + \frac{{1687}}{{60}}$ \\
& Refractive index & 1.44 \\
& Anisotropy & 0.8\\
\hline
\end{tabular}}
\end{center}
\end{table*}

For epidermis layer, dermis layer, blood, and hypodermis layer, we assigned them the respective optical properties according to Table 1. 
Some formulas or random values were determined according to Ciaran Bench et al’s work \cite{bench2020towards} and Jacques et al’s work \cite{jacques2013optical}.

Finally, 3890 different 3D skin tissue phantoms with specific optical properties were generated which are well prepared for the following optical and acoustic simulations.

\section{OPTICAL SIMULATION}

After generating the 3D human skin dataset, the optical fluence simulation was deployed based on the 3D Monte Carlo method. 
Motivated by Tang’s work \cite{tang20203d}, we use the MATLAB package MCXLAB released by Qianqian Fang \cite{fang2009monte} to simulate the propagation path of the illuminated photons in the 3D skin tissue models. 

\begin{figure*}[!t]
\centering
\includegraphics[width=7.1in]{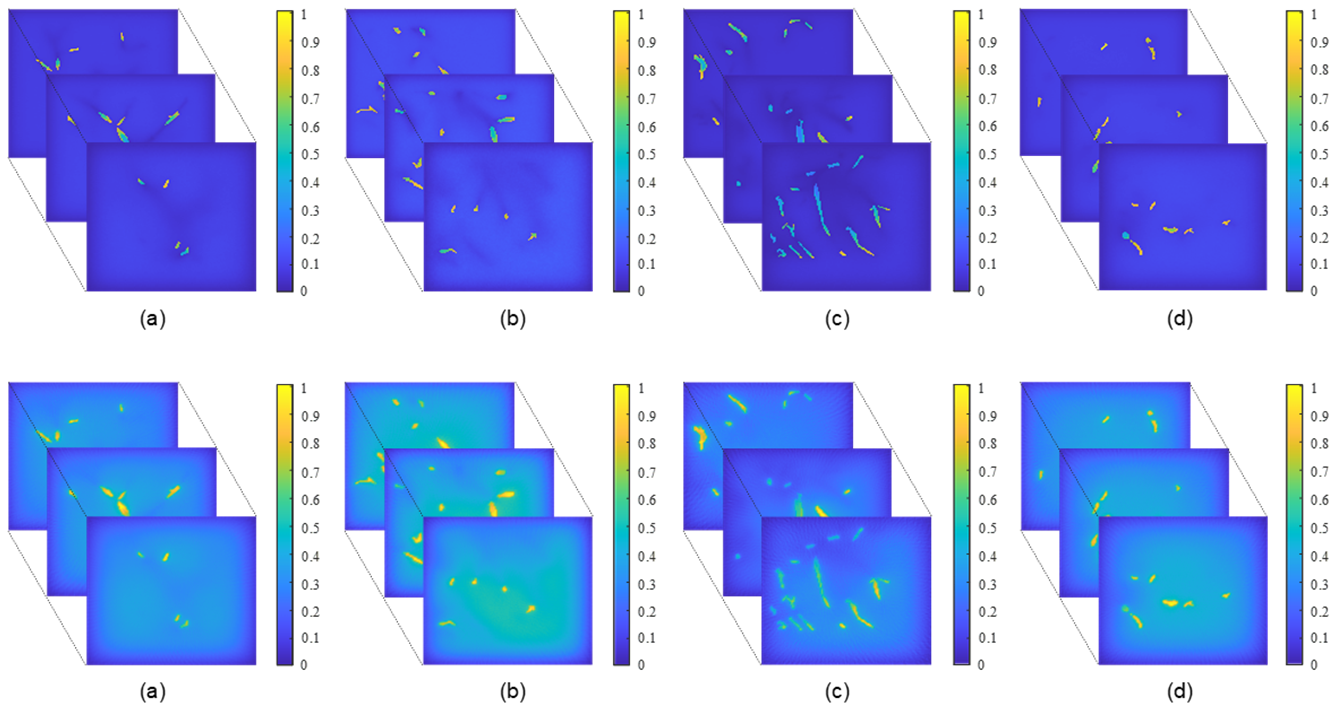}
\caption{Optical fluence slice map (the first row) and acoustic initial pressure slice map (the second row), the slice positions are chosen along the Z-axis which are 10, 20, 30 with respect to the total thickness of 64 voxels.}
\end{figure*}

A planar shape-like laser source pattern was designed to cover the whole surface of the epidermis. 
A billion photons were emitted for each optical simulation. The planar source was illuminated on the skin surface and the photons were absorbed by the four kinds of tissues which are epidermis, dermis, hypodermis, and blood vessel. According to Table 1, the optical fluence was computed and obtained. 
This simulation was run on a server with 4 GTX 1080Ti GPUs. It took us about 40 hours to simulate all these 3890 skin tissue models. 
We performed the optical absorption of the 3D skin tissue models at 16 different wavelengths from 750 nm to 900 nm with 10 nm interval. 
After about 26 days’ running on the server, we got 16 different sets of optical fluence maps for each model. 
The size of each 3D skin tissue model’s optical fluence map is about 3.4Mb. 
The optical fluence maps are shown in the first row of Fig. 4, where positions are chosen along the Z-axis which are voxel 10, 20, 30 with respect to the total thickness of 64 voxels.

\section{ACOUSTIC SIMULATION AND IMAGE RECONSTRUCTION}

By multiplying the simulated optical fluence and optical absorption coefficient map, the initial acoustic pressure distribution can be obtained, which is used for the following acoustic simulation. 
The PA generation and acoustic propagation simulation in the 3D skin tissue models were run based on the k-Wave toolbox developed by Bradley Treeby et al \cite{treeby2010k}. 
The simulation strategy in this work is to run the forward propagation process slice by slice, that is, for every 3D skin tissue model, we simulated the acoustic forward model and image reconstruction slice by slice. 
Then by stacking these reconstructed PA images together, we can obtain the reconstructed 3D PA initial pressure maps. 
The reconstruction method used here is universal back-projection proposed by Minghua Xu et al \cite{xu2005universal}, which turns out to be faster and provide better reconstruction image quality than conventional time-reversal and delay-and-sum methods. 
The acoustic sensor array was set to be circular, and the number of the sensor position was 180 to cover the whole region of the skin tissue model. 
In other words, we did not use a direct 3D acoustic simulation which is time-consuming, instead, a 2D modality was adopted. 
In fact, the slice-by-slice simulation setup seems like a real-scenario cylindrical scan. 
Stacking these 64 slices back together is reasonable. 
In the second row of Fig. 4, we showed four samples out of 3890 3D skin tissue models, in which the positions are chosen along the Z-axis that are voxel 10, 20, 30 with respect to the total thickness of 64 voxels.

\section{CONCLUSIONS}

In this paper, we basically and mainly talk about how to make a complex and real enough 3D skin vascular model to simulate the physical process of the PA effect. 
By embedding the CT lung vessels into some fixed-size cube shell, 3890 skin vascular tissue models were generated. 
And for every model, we assigned them a unique optical property according to the findings of the existing research. 
In the 3D Monte Carlo optical simulation section, we directly run the 3D program to obtain the optical fluence maps. In the acoustic propagation and image reconstruction section, we adopted a 2D strategy, which is run slice by slice and then stacked them together to obtain the 3D PA initial pressure distribution. 
It turns out that our 3D skin vascular model dataset is complex enough and close to the real scenario, which could be popularized. 
In the future, this dataset can be used for lots of research work in PAT, such as blood oxygen saturation quantification, PA image classification, registration and reconstruction.

\addtolength{\textheight}{-12cm}   




\section{ACKNOWLEDGMENT}
This research was funded by Start-up grant of ShanghaiTech University (F-0203-17-004), Natural Science Foundation of Shanghai (18ZR1425000), and National Natural Science Foundation of China (61805139). The authors declare no conflicts of interest.

\bibliographystyle{IEEEbib}
\bibliography{strings,refs}

\end{document}